\documentclass[aps,prl,twocolumn,superscriptaddress,showpacs,amsmath,amssymb,longbibliography]{revtex4-1}
\usepackage{graphicx}
\usepackage[breaklinks]{hyperref}
\usepackage[english]{babel}
\usepackage{amsmath}
\usepackage{xcolor}
\usepackage{ulem}

\usepackage{color}
\definecolor{nred} {RGB}{224,0,0}
\definecolor{nblue} {RGB}{28,130,185}
\definecolor{dgreen} {RGB}{78,138,21}

\usepackage[sort&compress]{natbib}
\begin{document}


\title{Instability of subdiffusive spin dynamics in strongly disordered Hubbard chain}

\author{M. {\'S}roda }
\affiliation{Department of Theoretical Physics, Faculty of Fundamental Problems of Technology, Wroc\l aw University of Science and Technology, 
50-370 Wroc\l aw, Poland}

\author{P. Prelov\v sek}
\affiliation{J. Stefan Institute, SI-1000 Ljubljana, Slovenia }
\affiliation{Faculty of Mathematics and Physics, University of Ljubljana, SI-1000 Ljubljana, Slovenia }

\author{M. Mierzejewski}
\affiliation{Department of Theoretical Physics, Faculty of Fundamental Problems of Technology, Wroc\l aw University of Science and Technology, 
50-370 Wroc\l aw, Poland}

\begin{abstract}
We study spin transport in a Hubbard chain with strong, random, on--site potential and with spin--dependent hopping integrals, $t_{\sigma}$. For the the SU(2) symmetric case,  $t_{\uparrow} = t_{\downarrow}$, such model exhibits only partial many-body localization with localized charge  and (delocalized) subdiffusive spin excitations. Here, we demonstrate that breaking the SU(2) symmetry by
 even weak spin--asymmetry,  $t_{\uparrow} \ne t_{\downarrow}$,   localizes spins and restores full many-body localization. To this end we derive an effective spin model, where
 the spin subdiffusion is shown to be destroyed by arbitrarily weak $t_{\uparrow} \ne t_{\downarrow}$.  Instability of the spin subdiffusion originates from an interplay between
 random effective fields and  singularly distributed random exchange interactions. 
\end{abstract}

\maketitle

{\it Introduction--}
Many---body localization (MBL) is one of the most challenging phenomena in condensed matter physics  \cite{basko06,oganesyan07}  which has recently stimulated intensive theoretical and experimental studies concerning the low--dimensional strongly disordered 
many-body systems.  Theoretical studies considered and identified MBL predominantly in the chains of interacting spinless  fermions (or in equivalent Heisenberg--like spin models) \cite{monthus10,luitz15,ZZZ5_4,Ponte2015,lazarides15,vasseur15a,serbyn2014a, pekker2014,torres15,torres16,laumann2015,huse14,gopal17,Hauschild_2016,herbrych13,imbrie16,steinigeweg16}. 
Among distinctive properties of such disordered systems, there is  absence of thermalization in the MBL phase \cite{znidaric08,bardarson12,kjall14,serbyn15,luitz16,serbyn13_1,bera15,altman15,agarwal15,gopal15,znidaric16,mierzejewski2016,lev14,lev15,barisic16,bonca17,bordia2017_1,zakrzewski16,protopopov2018,sankar2018,zakrzewski2018} and, moreover, 
unusually slow equilibration also beyond the boundaries of the MBL regime. In particular, the  subdiffusive dynamics has been found in  several  one-dimensional models for  moderate disorder and  has been identified as a precursor to  MBL \cite{luitz2016prl,luitz116,znidaric16,gopal17,kozarzewski18,prelovsek217,new_karrasch,prelovsek2018a}.

The qualitative features of MBL  have  been confirmed  in several experimental  studies of cold--fermion lattice systems \cite{kondov15,schreiber15,choi16,bordia16,bordia2017_1,smith2016} which, however, address the physics of the disordered Hubbard model
with both density (charge) and spin degrees of freedom.  The remaining SU(2) spin--symmetry of the latter models poses essential limitations 
to the existence of  the full MBL \cite{Chandran2014,potter16,prelovsek16,proto2017,friedman2017}. While the charge degrees of freedom appear
to be localized for sufficiently strong disorder, the spin degrees remain delocalized and undergo a subdiffusive dynamics \cite{prelovsek16,zakrzewski2018, kozarzewski18}.  
This implies that only partial (charge) MBL may occur in the SU(2) symmetric Hubbard chains. This scenario
is consistent with the number of  local integrals of motion \cite{mierzejewski2018} which stays well below the value expected for systems with full MBL. 
Moreover, one cannot exclude  that coupling of localized charges and delocalized spins will eventually delocalize also the charge degrees of freedom 
\cite{proto2018}, even if the latter delocalization will happen at exceedingly long time--scales.  

 
In this paper we reconsider the problem of full/partial MBL   and demonstrate that  in strongly 
disordered Hubbard model, the subdiffusive (but ergodic) spin dynamics is unstable against even weak perturbations that break
the SU(2) spin symmetry. In particular, we consider Hubbard chain with random on--site potential and with anisotropic (spin--dependent) hopping integrals, $t_{\sigma}$.
We study  
the long--time ($t \to\infty$) behavior of  local spin--spin correlations,  $C_0=\lim_{t\rightarrow \infty} \langle S^z_i(t) S^z_i(0) \rangle$, 
representing local spin stiffness being also an indicator of nonergodicity. While $C_0=0$ in the SU(2) symmetric case 
 ($t_{\uparrow}=t_{\downarrow}$), in agreement with a subdiffusive dynamics, it is shown to be non-zero even for very weak  hopping asymmetry. For asymmetric 
 hopping, it  exhibits a power--law 
 dependence $C_0 \propto |t_{\uparrow}-t_{\downarrow} |^{\gamma}$ indicating that full MBL is restored.   
In order to explain the instability of the spin subdiffusion, we derive  an effective (squeezed) model which describes the 
dynamics of  spin excitations. The latter  model takes the form of the  Heisenberg chain with random exchange interactions but 
also with random local magnetic fields.  The interplay between  random spin interactions  (with a singular distribution \cite{kozarzewski18}) 
and random fields appears to be responsible for  the spin localization and  restoration of full MBL for arbitrarily small difference $t_{\uparrow}-t_{\downarrow}$.
The numerical results for the Hubbard model in this regime confirm the simplified model and a general scenario.
  
{\it Model and method--}  We study a disordered Hubbard chain  
\begin{eqnarray}
H &=& H_0 + U \sum_i n_{i\uparrow}n_{i\downarrow},  \label{hubb} \\
H_0&=&   -\sum_{i,\sigma} t_{\sigma} c^{\dagger}_{i \sigma} c_{i+1 \sigma} +{\rm H.c.} +\sum_i \epsilon_i (n_{i \uparrow}+n_{i \downarrow}),  \label{h0}
\end{eqnarray} 
where $c^{\dagger}_{i\sigma} $ creates  a fermion with spin $\sigma$ at site $i$, $n_{i \sigma}=c^{\dagger}_{i\sigma}c_{i\sigma}$ and   
the  disorder enters only via random potentials,  $ \epsilon_i$, which are  uniformly distributed in $ [-W,W]$. 
The spin asymmetry is introduced via  hopping integrals, where we adopt   $ t_{\uparrow}=1$  as the energy unit, while $t_{\downarrow} \le 1$.

As it follows from the experimental \cite{schreiber15}  and theoretical \cite{prelovsek16,zakrzewski2018} studies, the charge dynamics  in the Hubbard chain (\ref{hubb}) 
is frozen  for sufficiently strong disorder, $W \gg 1$. Therefore, it is useful and sufficient to derive a squeezed  model which involves only 
spin degrees of freedom.  To this end we diagonalize the single--particle Hamiltonian, 
$H_0=\sum_{a ,\sigma} \varepsilon_{a\sigma}  c^{\dagger}_{a\sigma} c_{a\sigma} $, where  $c^{\dagger}_{a \sigma}=\sum_i \phi_{i a \sigma }   c^{\dagger}_{i \sigma} $ creates a fermion in the Anderson state 
and we take all $ \phi_{i a \sigma }$ as real. We consider 
only strong disorder $W\gtrapprox 4$, when the single--particle localization length is very short, $\lambda <1$. 
For convenience, the Anderson states are sorted according to the maxima of  $ |\phi_{i a \sigma } |$ in  real--space so that
$\phi_{i a \uparrow}$  and $\phi_{i a \downarrow}$ are centered in the vicinity of the same lattice site, $i$,  despite $\phi_{i a \uparrow}  \ne \phi_{i a \downarrow}$. 
Consequently the quantum number  $a$  marks positions of the Anderson states in real space.
 
 In order to obtain the squeezed spin model, we rewrite the Hubbard term in Eq. ($\ref{hubb}$) using the Anderson basis \cite{kozarzewski18}. 
 In view of the frozen charge dynamics, we keep  only terms which do not alter the occupancy of the Anderson states, i.e., we keep terms commuting 
 with $n_a= n_{a \uparrow}+n_{a \downarrow}$.  Then we can rewrite the effective
 Hamiltonian using the spin operators, $S^z_{a}=\frac{1}{2}(n_{a \uparrow}-n_{a \downarrow} )$, $S^+_{a}= c^{\dagger}_{a \uparrow}c_{a \downarrow} $ and  $S^-_{a}= c^{\dagger}_{a \downarrow}c_{a \uparrow} $,
  \begin{eqnarray}
 H \simeq -  \sum_{a<b} \left[ J^z_{ab} S^z_a S^z_b + \frac{J^{\perp}_{ab}}{2}(S^+_a S^-_b+S^-_a S^+_b)\right]+\sum_a h_a S^z_a,  \nonumber \\
 \label{heff}
 \end{eqnarray} 
 where 
  \begin{eqnarray}
 J^z_{ab}&=& U \sum_i \left[ (\phi_{ia \uparrow} \phi_{ib \downarrow})^2+ (\phi_{ia \downarrow} \phi_{ib \uparrow} )^2 \right],  \label{jz}  \\ 
 J^{\perp}_{ab} &=& 2 U \sum_i   \phi_{ia \uparrow} \phi_{ib \downarrow} \; \phi_{ia \downarrow} \phi_{ib \uparrow},  \label{jp} \\
 h_a &=& \Delta \varepsilon_{a} + \frac{U}{2}  \sum_{b\ne a,i}  n_{b} 
 \left[(\phi_{ia \uparrow} \phi_{ib \downarrow})^2- (\phi_{ia \downarrow} \phi_{ib \uparrow} )^2 \right], \label{ha}
 \end{eqnarray}
 with $\Delta \varepsilon_a= \varepsilon_{a\uparrow} - \varepsilon_{a\downarrow}$. On the one hand, starting from the SU(2) symmetric Hubbard chain ($\phi_{ia \uparrow} =  \phi_{ia \downarrow}$) one obtains a SU(2) symmetric model with $J^z_{ab}= J^{\perp}_{ab}$ and $h_a=0$, where spins  have been shown to be delocalized  and the spin transport is subdiffusive \cite{kozarzewski18}.  For  $\phi_{ia \uparrow} \ne  \phi_{ia \downarrow}$ the effective model takes the form of an easy--axis XXZ model with random $J^z_{ab} \ge |J^{\perp}_{ab}|$ but also with random fields $h_a$. Due to the latter interaction, Eq.~(\ref{heff}) resembles the canonical model studied in the  context of MBL \cite{monthus10,luitz15,ZZZ5_4,Ponte2015,lazarides15,vasseur15a,serbyn2014a, pekker2014,torres15,torres16,laumann2015,huse14,gopal17,Hauschild_2016,herbrych13,imbrie16,steinigeweg16}.
However, an essential difference in Eq. (\ref{heff})  emerges from random interactions $J_{ab}^z, J_{ab}^\perp$ with singular distributions, as shown later on.
Hence,  dynamical properties cannot be simply deduced from previous studies of the standard model.
 
 In order to study numerically the spin dynamics, we first generate random $\epsilon_i$ in Eq. (\ref{h0})  and diagonalize
 $H_0$ for a chain of $L$ sites. Then, we randomly choose $N$ Anderson states occupied by fermions ($N/2$ for each spin projection).  
 Doubly occupied states $|a\rangle$  are spin singlets and do not contribute to the Hamiltonian  (\ref{heff}).   
 Consequently, the squeezed model contains (on average) only  $\tilde{L}  \sim N-N^2/2L$ singly--occupied states $|a\rangle$.  
 Note  that the average distance  between fermions occupying these states  is $L/\tilde{L} \ge 2 \gg \lambda $,  
 even for the half--filled Hubbard model,  $\bar{n}=N/L=1$. Moreover, overlaps of the wave--functions in Eqs. (\ref{jz}-\ref{ha}) decay exponentially with the real--space distance 
 between the Anderson states, hence we consider  interactions only between nearest neighbors $b=a \pm 1$.
  
  \begin{figure} 
 \includegraphics[width=\columnwidth]{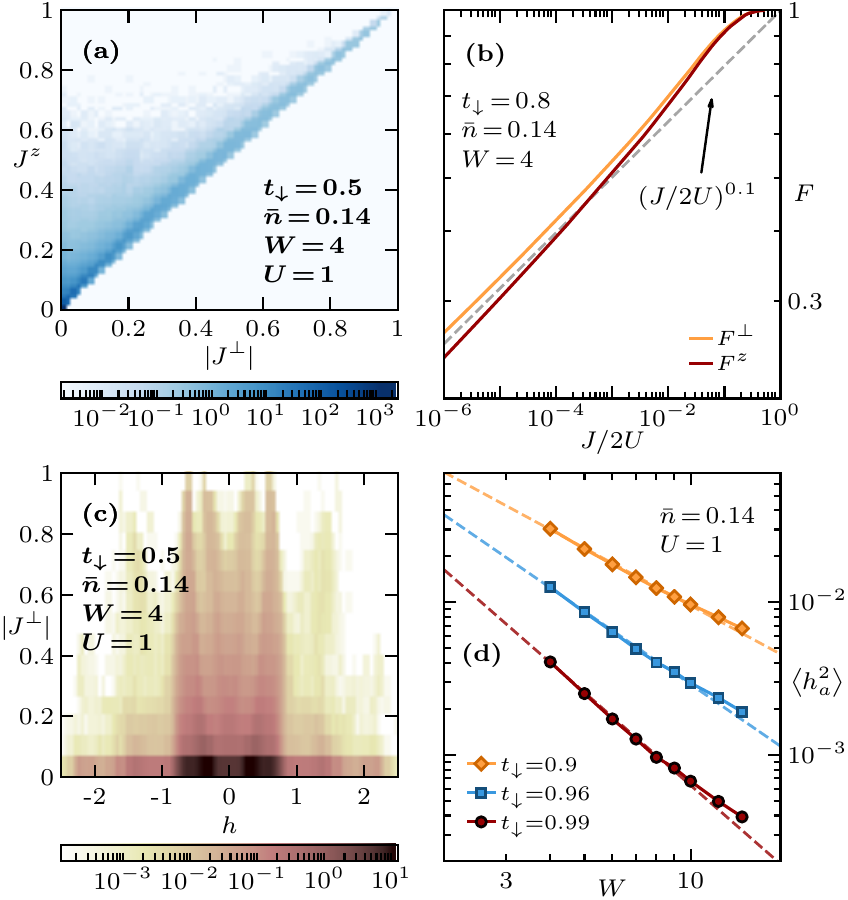}
\caption{  a) and c) Joint probability densities, 
 $p(|J^{\perp}|, J^z)$ and $p(|J^{\perp}|, h)$, respectively, for filling $\bar{n}=N/L=0.14$. b)
Cumulative distribution functions $F^z$, $F^{\perp}$ compared with analytical results for SU(2) symmetric case, $F^0$ (dashed line). d) 
Variance of the random field $\langle h^2_a \rangle$ in the squeezed model  as a function of disorder $W$. }   
\label{fig1}
\end{figure} 

 Let us first consider statistical properties of parameters $J^z_{a}, J^{\perp}_{a}, h_a$ as they  occur in the squeezed model (\ref{heff}). 
Fig. \ref{fig1}a   shows the joint probability density, $p(|J^{\perp}|, J^z)$. One may observe that $J^z$ and  $J^{\perp}$ are strongly correlated with each other and Fig. \ref{fig1}a  reveals a clear maximum at $J^z \simeq |J^{\perp}|$.  Moreover, the probability 
densities, $f^z( J^z)$ and   $f^{\perp}(|J^{\perp}|)$, are rather insensitive to a modest difference $t_{\uparrow}-t_{\downarrow}$.   This can be observed from the cumulative distribution functions $F^z(J)=\int_0^{J} \mathrm{d} J'  f^z(J') $  and $F^{\perp}(J)=\int_0^{J} \mathrm{d} J'  f^{\perp}(|J'|) $ shown in Fig.\ref{fig1}b for $t_{\downarrow}=0.8$.
They are quite close to the distribution $F^0(J)=(J/2U)^{\tilde{\lambda}}$ (also plotted in Fig.\ref{fig1}b) with  $\tilde{\lambda}=\lambda \tilde{L}/L$ which describes the 
 distribution of  $J$ in the SU(2) symmetric model \cite{kozarzewski18}. The similarity of distributions is important, since they are singular also for the asymmetric model (provided that 
$\tilde \lambda <1$). We stress that  probability for $J^\perp_a=0$ vanishes. The latter would induce a trivial spin localization via cutting the chain into disconnected parts.  
Again, $\tilde{\lambda}$ is the essential parameter which in the SU(2) symmetric case governs the  subdiffusive dynamics, $ \langle S^z_a(t) S^z_a(0) \rangle \propto t^{-\tilde{\lambda}/(1+\tilde{\lambda})}$.
In all considered cases we also find $\tilde{\lambda} >0$, i.e., random $J$ alone is insufficient to cause spin localization, at finite temperatures $T>0$.   
   
Fig.\ref{fig1}d shows the strength of  random magnetic field $h_a$, i.e. the variance $ \langle h_a^2 \rangle$. 
 This quantity shows a power--law decrease with the disorder strength $W$ for arbitrary $t_{\uparrow} \ne t_{\downarrow}$.
 Counterintuitively,  strongly  disordered Hubbard model maps onto the spin chain with random fields  which are too weak to cause an efficient spin localization.
 The essential physical mechanism behind the onset of  spin--localization can be observed in  Fig.\ref{fig1}c which shows the joint probability density 
 $p(|J^{\perp}|, h)$. When compared to Fig.\ref{fig1}a, correlations between $J^{\perp}_a$ and  $h_a$ seem insignificant. 
Therefore, there is  quite high probability for regions with  large ratio $h_a/J^\perp_a$, which in the following are shown to be essential for the spin localization.
    
\begin{figure} 
 \includegraphics[width=\columnwidth]{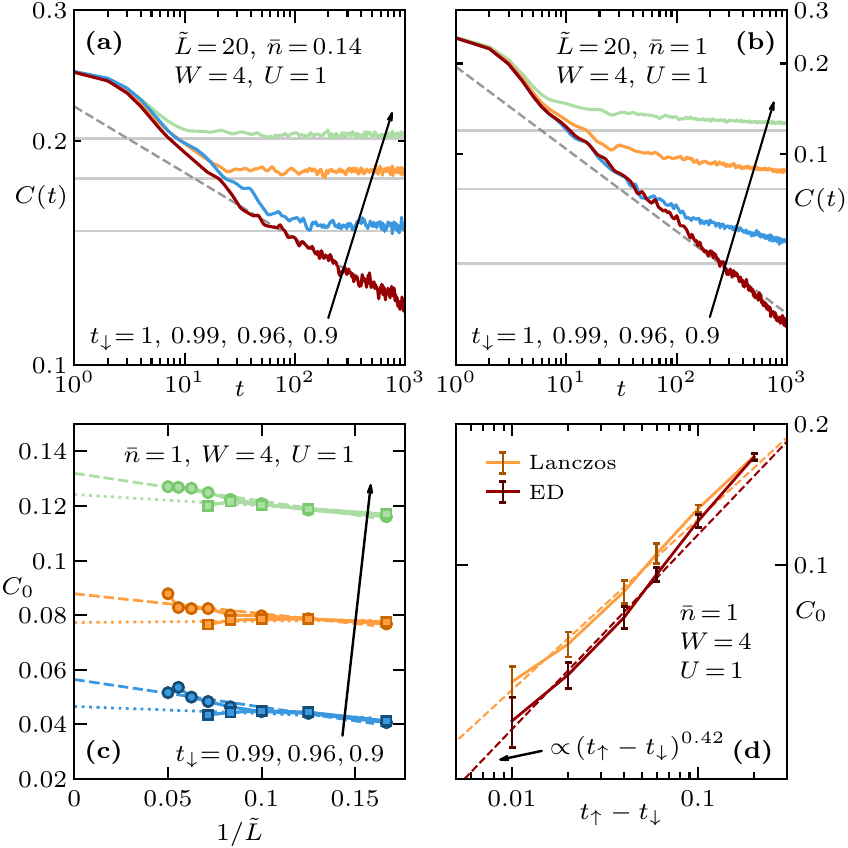}
\caption{a) and b): Local spin--spin correlation function, $C(t)$,  for fillings $\bar n=0.14$  and $\bar n=1$, respectively. Results  for fixed disorder $W=4$ and 
various $t_{\downarrow}$  are obtained from the Lanczos method. Horizontal lines show stiffnesses $C_0$  obtained from ED for $\tilde{L}= 14 $.  The dashed lines show approximate power-law for SU(2) symmetric case, $C(t) \propto t^{-\tilde{\lambda}/(1+\tilde{\lambda}) }$.
c) Spin stiffnesses $C_0$ obtained from Lanczos  method (circles) and from ED (squares)  extrapolated linearly in $1/\tilde{L} \to 0$, and 
d)  corresponding extrapolated values $C_0$ vs. $t_{\uparrow}-t_{\downarrow}$.  Error bars in d) show uncertainty of the finite-size scaling.}
\label{fig2}
\end{figure} 
\begin{figure} 
\includegraphics[width=\columnwidth]{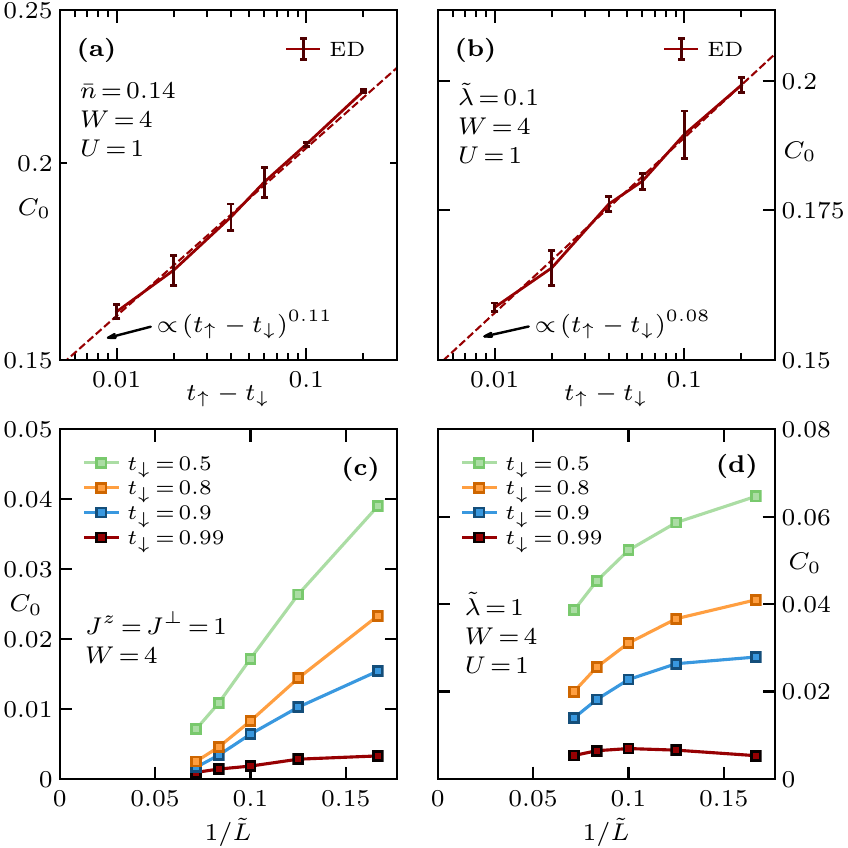}
\caption{a)  Spin stiffnesses, $C_0$,  as in Fig.\ref{fig2}d but for smaller filling $\bar{n}=0.14$, b) corresponds to a), but for simplified squeezed model,
while c)  and d)  refer to  Fig.\ref{fig2}c but for simplified model with
constant $J$ (c) and uniformly distributed  random $J$ (d). Note logarithmic scales on both axes in a) and b).}   
\label{fig3}
\end{figure} 
    
 Figs. \ref{fig2}a, b show the central result of this {\it work}: the local spin--spin correlation function for the effective spin model
\begin{equation} 
C(t)= \langle \psi | S^z_a(t) S^z_a(0) | \psi \rangle_{\rm ave}. \label{cs}
\end{equation}	  
We have calculated $C(t)$   taking into account parameters from the original Hubbard model in accordance with Eqs. (\ref{jz})--(\ref{ha}).
Then, numerical results have been averaged over random $\epsilon_i$, as well as over random choice of singly occupied Anderson states $ |a \rangle $.
Averaging over $| \psi \rangle$ in the corresponding  squeezed model has been carried out at high temperature, $T \to \infty$.
We stress that results still depend on the filling $\bar n$ of the Hubbard chain and on  the size  of the squeezed chain $\tilde L$.

As follows from Fig. \ref{fig1}b, there is a quite elevated probability for finding  weak--links \cite{agarwal15,bordia2017_1} with small $J^{\perp}_a$, which may result in long--lasting but still transient phenomena.
In order to rule out such transient effects we have used two complementary numerical methods and verified their consistency. 
Namely, data for $\tilde L \sim 20$ and times $t \lessapprox 10^3$ are obtained via the time-dependent Lanczos method \cite{park86}, whereas longer times 
but smaller systems  $\tilde L \leq 14$ are studied by exact diagonalization (ED).  We have carried out averaging over $10^3$
and $10^4$ realizations of disorder for the former and the latter methods, respectively.
For the symmetric case, $t_{\downarrow}=1$,  $C(t)$  
shows unrestricted power--law decay in time corresponding to ergodic but subdiffusive behavior. However, it saturates even  for   very weak asymmetry 
$t_{\uparrow}-t_{\downarrow} \sim 10^{-2}$ marking the onset of nonergodicity and spin localization. It holds true not only  at low--filling ($\bar n \ll 1$)
shown in  Fig.\ref{fig2}a  but also for parameters corresponding to the half-filled Hubbard chain ($\bar n=1$)  in Fig.\ref{fig2}b.  
Fig.\ref{fig2}c presents a finite size scaling of the spin stiffness $C_0=C (t\rightarrow \infty)$ vs. $1/\tilde L$. More precisely, the circles show result
obtained from  the Lanczos method for $t=10^3$,  whereas squares show  $C_0$ obtained via ED.  It appears, that both approaches yield very similar results 
for the extrapolated stiffness $C_0$, as presented in Figs. \ref{fig2}d and \ref{fig3}a for fillings $\bar  n=1 $ and  $\bar n \simeq 0.14 $, respectively.
Finally, extrapolated stiffness shows a power-law dependence on the asymmetry parameter,  i.e. $C_0 \propto (t_{\uparrow}-t_{\downarrow})^{\gamma}$ and 
$\gamma$ is of the order of $\tilde{\lambda}$. Consistently with previous considerations \cite{kozarzewski18}, for symmetric case 
$t_{\downarrow}=1$ we get ergodic behavior with $C_0=0$, but  with a subdiffusive dynamics provided that $\tilde \lambda <1$.
 

It follows from  Fig.\ref{fig1}b that $f^{\perp}(J) \simeq f^z(J) \simeq f^0(J)= \tilde{\lambda} (J/2U )^{\tilde{\lambda}-1}$ for $J \le 2U$. The power--law distribution 
has an integrable singularity at $J=0$ provided that $0< \tilde{\lambda} < 1$. Namely, the singularity occurs when the average 
distance between singly occupied Anderson states $L/\tilde{L}$ is larger than the single--particle localization length $\lambda$. 
One may further simplify the squeezed model,  Eqs. (\ref{jz})-(\ref{ha}), assuming that $J^{\perp}_a=J^z_a=J_a$ are random variables with a distribution function 
$f^0(J)$ and that random $J_a$ is uncorrelated with field  $h_a \simeq  \Delta \varepsilon_{a}$.   Figs. \ref{fig3}a and \ref{fig3}b show  the comparison of $C_0$, 
obtained for the complete and the simplified squeezed model, respectively. One may observe that the simplified version indeed maintains the essential 
properties of the more general version. Moreover, the simplified version allows to study  regimes which cannot be derived from the Hubbard model within 
our approach, e.g.,  when $W$ is small or $U$ is too large.  In particular,  Fig.\ref{fig3}d shows the same
results as Fig.\ref{fig2}c but obtained for $\tilde{\lambda}=1$, i.e.,  for a uniform (nonsingular) distribution of random $J$.  Despite the presence of the 
random fields, $h_a$, spins remain delocalized for non--singular $f(J)$. Absence of spin localization can  also be observed in Fig.\ref{fig3}c that shows 
results for random $h_a$ but with uniform $J^z_a=J^{\perp}_a=1$.  Concluding this part, we can therefore stress that the instability of the spin 
subdiffusion and the onset of spin localization originate  in the squeezed model from a coexistence of  random fields $h_a$ and the singular
distribution of random $J^{\perp}_a \simeq J^{z}_a$.

\begin{figure} 
 \includegraphics[width=\columnwidth]{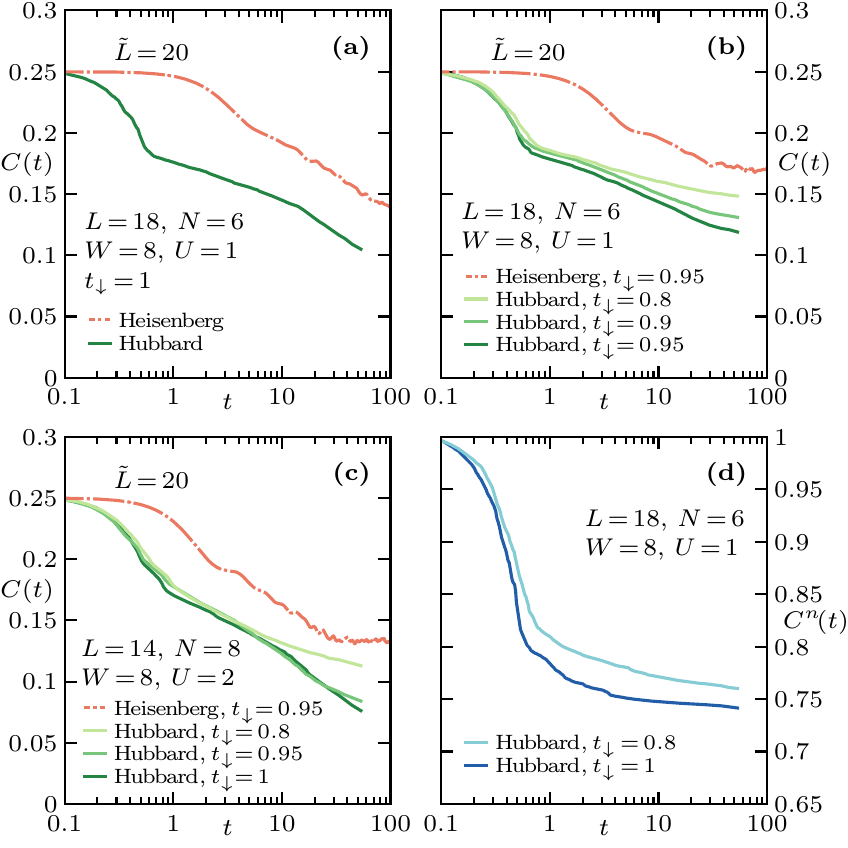}
\caption{$C(t)$ obtained for the original disordered Hubbard model, compared with results for the corresponding squeezed (Heisenberg) model
at fixed disorder $W=8$: a) for $\bar n=1/3$ and SU(2) symmetric case, and b) for different assymmetries $t_\downarrow < 1$,  and c) filling $\bar{n}=8/14 = 0.57$ closer 
to quarter-filling.  d) Charge--charge correlation function $C^n(t)$  for the  Hubbard model.}
\label{fig4}
\end{figure} 

As a final numerical support for our approach, we compare the local spin--spin correlation functions obtained from the squeezed model  and 
directly from the disordered full Hubbard chain, where $ C(t)= \langle S^z_i(t) S^z_i(0) \rangle$ is calculated at $T \to \infty$ 
via the microcanonical Lanczos method  \cite{prelovsek16,kozarzewski18}.
Since the local correlations  in the Hubbard model are defined in terms of the Wannier states $|i\rangle$, one can expect quantitative but not qualitative differences 
at $t \to \infty$.  A comparison is shown in Figs. \ref{fig4}a, b, c for various parameters.   
Due to much larger Hilbert space,  results for the Hubbard chains are obtained for rather limited system sizes 
$L \leq 18$, low fillings $\bar n \le 0.6$  and time--windows $t \leq 100$.  Both $C(t)$  reveal decay in time for
the SU(2) symmetric case (Fig.\ref{fig4}a)  and saturation for $t_{\downarrow} \neq t_{\uparrow}$ as shown in  Figs. \ref{fig4}b and \ref{fig4}c.  
The best agreement between the models is expected to show up  for modest $U < 4$,  large $W > 4$ (small $\lambda$) and low filling $\bar n \ll 1$ 
(large distance between spins). It is still satisfactorily close to quarter--filling, $\bar n \simeq 1/2$, (Fig. \ref{fig4}c), i.e. for the case studied experimentally \cite{schreiber15,bordia16}.

While the deviation between both
models at  $ t \lessapprox  1$ is not surprising, it is useful to explain its origin. To this end, for the Hubbard chain we have calculated  also a (normalized) charge--charge correlation function,
$C^n(t)= \langle (n_i(t)-\bar n)  (n_i-\bar n)\rangle/\bar n^2$, shown in Fig. \ref{fig4}d. We note, that in the short--time regime 
the charge is redistributed over the Anderson states. This feature is missing in the squeezed model and  is
responsible also for the short--time deviations visible in the spin dynamics of both $C(t)$.
  
{\it Conclusions. --}  We have studied how the spin dynamics in the disordered Hubbard chain depends on spin--dependent hopping that breaks the SU(2) symmetry. 
To this end we have derived an effective spin model assuming that the disorder strength is the largest energy scale (i.e., interaction is weak)
and the single--particle localization length is much smaller than the average distance between singly occupied sites. Results obtained for the squeezed model show that the subdiffusive spin dynamics occurs 
only for strictly  SU(2) symmetric  system whereas arbitrary  $t_{\downarrow} \ne t_{\uparrow}$ localizes spins and restores full MBL in the original Hubbard  model.
Instability of the subdiffusive dynamics originates from the interplay between two specific properties of the squeezed model: weak random magnetic field and random $J$ with a distribution function
that is singular at $J=0$.   Despite rather obvious numerical limitations, results obtained for the Hubbard chain qualitatively agree with those for the squeezed Hamiltonian.
There are nevertheless  open questions. It is unclear whether the instability of subdiffusive spin dynamics is restricted to the regime where we can reliably 
derive the squeezed spin model.  Moreover, we have considered only a specific breaking of SU(2) symmetry, and it is a pertinent question 
whether this instability holds for arbitrary perturbation that breaks the latter symmetry.

\acknowledgments  
 M. {\'S}. and M.M. acknowledge  support by the National Science Centre, Poland via project 2016/23/B/ST3/00647.
 P.P. is supported by the program P1-0044 and project N1-0088 of the Slovenian Research Agency.

\bibliography{ref_mbl}

\end{document}